\documentclass[conference,10pt]{IEEEtran}
\usepackage{epsfig,rotating,setspace,latexsym,amsmath,epsf,amssymb,amsfonts,bm,theorem,subfigure,epstopdf}
\usepackage{cite,authblk}
\usepackage{url}
\usepackage{bbm}
\usepackage{algorithm,algpseudocode}

\IEEEoverridecommandlockouts

\allowdisplaybreaks
\usepackage{xcolor}
\usepackage{graphicx}

\newcommand{\defeq}{\ensuremath{\triangleq}}

\begin{document}

\title{Age-Based Coded Computation for\\
Bias Reduction in Distributed Learning
\thanks{This work was supported in part by the Marie Sklodowska-Curie Action SCAVENGE (grant agreement no. 675891), and by the European Research Council (ERC) Starting Grant BEACON (grant agreement no. 677854).}
}

	\author[1]{Emre Ozfatura}
	\author[2]{Baturalp Buyukates}
	\author[1]{Deniz Gunduz}
	\author[2]{Sennur Ulukus}
	\affil[1]{\normalsize Department of Electrical and Electronic Engineering, Imperial College London, UK}
	\affil[2]{\normalsize Department of Electrical and Computer Engineering, University of Maryland, MD, USA}
	

\maketitle
\begin{abstract}
Coded computation can be used to speed up distributed learning in the presence of straggling workers. Partial recovery of the gradient vector can further reduce the computation time at each iteration; however, this can result in biased estimators, which may slow down convergence, or even cause divergence. Estimator bias will be particularly prevalent when the straggling behavior is correlated over time, which results in the gradient estimators being dominated by a few fast servers. To mitigate biased estimators, we design a \emph{timely} dynamic encoding framework for partial recovery that includes an ordering operator that changes the codewords and computation orders at workers over time. To regulate the recovery frequencies, we adopt an \emph{age} metric in the design of the dynamic encoding scheme. We show through numerical results that the proposed dynamic encoding strategy increases the timeliness of the recovered computations, which as a result, reduces the bias in model updates, and accelerates the convergence compared to the conventional static partial recovery schemes.
\end{abstract}

\section{Introduction}
One of the main factors behind the success of machine learning algorithms is the availability of large datasets for training. However, as datasets become ever larger, the required computation becomes impossible to execute in a single machine within a reasonable time frame. This computational bottleneck can be overcome by  distributed learning across multiple machines,  called {\em workers}. \\
\indent Gradient descent (GD) is the most common approach in supervised learning, and can be easily distributed. By employing a {\em parameter server} (PS) type framework \cite{PSGD4}, the dataset can be divided among workers, and at each iteration, workers compute  gradients based on their local data, which can be aggregated by the PS. However, slow, so-called {\em straggling}, workers are the Achilles heel of distributed GD (DGD) since the PS has to wait for all the workers to complete an iteration. A wide range of straggler-mitigation strategies have been proposed in recent years \cite{UCCT.1, UCCT.3, UCCT.4, UCCT.6, UCCT.7, CC.1, CC.2, CC.3, CC.4, CC.5, Yang19, CC.6, CC.8, CC.11, CC.17, CC.18, CC.AG1, entropy}. The main notion is to introduce redundancy in the computations assigned to each worker, so that fast workers can compensate  for the stragglers.\\
\indent Most of the coded computation solutions for straggler mitigation suffers from two drawbacks: First, they allow each worker to send a single message per iteration, which results in the under-utilization of computational resources \cite{entropy}. Second, they recover the full gradient at each iteration, which may unnecessarily increase the average completion time of an iteration. Multi-message communication (MMC) strategy addresses the first drawback by allowing each worker to send multiple messages per-iteration, thus, seeking a balance between computation and communication latency \cite{CC.2,CC.5,CC.11,CC.13,CC.23,CC.33, UCUT.4,UCCT.6}. \cite{icassp-ext} addresses the second drawback by combining coded computation with partial recovery (CCPR) to provide a trade-off between the average completion time of an iteration and the  accuracy of the recovered gradient estimate.

If the straggling behavior is not independent and identically distributed over time and workers, which is often the case in practice, the gradient estimates recovered by the CCPR scheme become biased. For example, this may happen when a worker straggles over multiple iterations. Regulating the recovery frequency of the partial computations to make sure that each partial computation contributes to the model updates as equally as possible is critical to avoid biased updates. We use the age of information (AoI) metric to track the recovery frequency of partial computations. 

AoI has been proposed to quantify the data freshness over systems that involve time-sensitive information \cite{Kaul12a}. AoI studies aim to guarantee timely delivery of time-critical information to receivers. AoI has found applications in queueing and networks, scheduling and optimization, and reinforcement learning (see the survey in \cite{SunSurvey}). Recently, \cite{Buyukates19c} considered the age metric in a distributed computation system that handles time-sensitive computations, and \cite{Yang19a} introduced an age-based metric to quantify the staleness of each update in a federated learning system. In our work, we associate an age to each partial computation and use this age to track the time passed since the last time each partial computation has been recovered.

In this paper, we design a dynamic encoding framework for the CCPR scheme that includes a timely dynamic order operator to prevent biased updates, and improve the performance. The proposed scheme increases the timeliness of the recovered partial computations by changing the codewords and their computation order over time. To regulate the recovery frequencies, we use \textit{age of the partial computations} in the design of the dynamic order operator. We show by numerical experiments on a linear regression problem that the proposed dynamic encoding scheme increases the timeliness of the recovered computations, results in less biased model updates, and as a result, achieves better convergence performance compared to the conventional static encoding framework.
\begin{figure}
\centering
\includegraphics[scale=0.4]{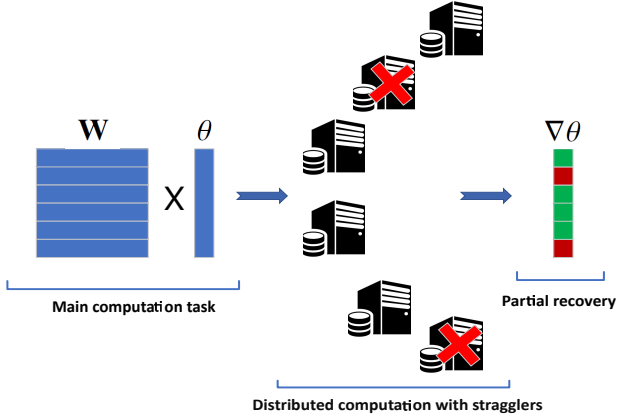}
\caption{Illustration of  partial recovery in a naive distributed computation scenario with 6 workers, 2 of which are stragglers.}
\label{erasure}
\vspace{-0.5cm}
\end{figure}
\section{System Model and Problem Formulation}
For completeness, we first present the coded computation framework and the CCPR scheme.
\subsection{DGD with Coded Computation }
We focus on the least-squares linear regression problem, where the loss function is the empirical mean squared error
\begin{equation}
L(\boldsymbol{\theta}) \triangleq \frac{1}{2N}\sum_{i=1}^{N}(y_{i}-\mathbf{x}_{i}^{T}\boldsymbol{\theta})^{2},
\label{loss}
\end{equation}
where $\mathbf{x}_{1},\ldots,\mathbf{x}_{N} \in \mathbb{R}^{d}$
are the data points with corresponding labels $y_{1},\ldots,y_{N} \in \mathbb{R}$, and $\boldsymbol{\theta}\in \mathbb{R}^{d}$ is the parameter vector. The optimal parameter vector can be obtained iteratively by using the gradient descent (GD) method
\begin{equation}
\boldsymbol{\theta}_{t+1}=\boldsymbol{\theta}_{t}-\eta_{t} \nabla_{\boldsymbol{\theta}} L(\boldsymbol{\theta}_{t}),
\end{equation}
 where $\eta_{t}$ is the learning rate and $\boldsymbol{\theta}_{t}$ is the parameter vector at the $t$th iteration. Gradient of the loss function in (\ref{loss}) is 
 \begin{equation}
 \nabla_{\boldsymbol{\theta}} L(\boldsymbol{\theta_t}) = \mathbf{X}^{T} \mathbf{X} \boldsymbol{\theta}_{t}-\mathbf{X}^{T}\mathbf{y}, 
 \label{grad_comp}
 \end{equation}
 where $\mathbf{X}=[\mathbf{x}_{1},\ldots,\mathbf{x}_{N}]^{T}$ and 
$\mathbf{y}=[y_{1},\ldots,y_{N}]^{T}$. In (\ref{grad_comp}), only $\boldsymbol{\theta}_{t}$ changes over iterations. Thus, the key computational task at each iteration is the  matrix-vector multiplication of $\mathbf{W}\boldsymbol{\theta}_{t}$, where $\mathbf{W}\defeq\mathbf{X}^{T}\mathbf{X}\in\mathbb{R}^{d\times d}$. To speed up GD, execution of this multiplication can be distributed across $K$ \textit{workers}, by simply dividing $\mathbf{W}$ into $K$ equal-size disjoint submatrices. However, under this naive approach, computation time is limited by the {\em straggling} workers \cite{CC.1}.

Coded computation is used to tolerate stragglers by encoding the data before it is distributed among workers to achieve certain redundancy. That is, with coded computation, redundant partial computations are created such that the result of the overall computation can be obtained from a subset of the partial computations. Thus, up to certain number of stragglers can be tolerated since the PS can recover the computation result without getting partial results from all workers. Many coded computation schemes, including MDS \cite{CC.1,CC.2,CC.8}, LDPC \cite{CC.3}, and rateless codes \cite{CC.23} and their various variants have been studied in the literature.

\subsection{Coded Computation with Partial Recovery (CCPR)}
In naive uncoded distributed computation for gradient computation, straggling workers result in erasures in the gradient vector as illustrated in Fig.~\ref{erasure}. The main motivation behind the coded computation schemes is to find the minimum number of responsible workers to guarantee the recovery of the gradient vector without any erasures. Alternatively, the CCPR scheme \cite{icassp-ext} allows erasures on the gradient vector to reduce the computation time while controlling the number of erasures to guarantee certain accuracy for the gradient estimate.

\indent To enable  partial recovery, we focus on a linear code structure such that $\mathbf{W}$ is initially divided into $K$ disjoint submatrices $\mathbf{W}_{1},\ldots,\mathbf{W}_{K}\in\mathbb{R}^{d/K\times d}$. Then, $r$ coded submatrices, $\tilde{\mathbf{W}}_{i,1},\ldots,\tilde{\mathbf{W}}_{i,r}$, are assigned to each worker $i$ for computation, where each coded matrix $\tilde{\mathbf{W}}_{i,j}$ is a linear combination of $K$ submatrices, i.e.,
\begin{equation}
\tilde{\mathbf{W}}_{i,j}=\sum_{k\in[K]}\alpha^{(i)}_{j,k}\mathbf{W}_{k}.
\end{equation}
 Following the initial encoding phase, at each iteration $t$, the $i$th worker performs the computations $\tilde{\mathbf{W}}_{i,1}\boldsymbol{\theta}_{t},\ldots,\tilde{\mathbf{W}}_{i,r}\boldsymbol{\theta}_{t}$ in the given order, and sends the results one by one as soon as they are completed. In the meantime, the PS collects coded computations from all the workers until it successfully recovers $(1-q) \times 100$ percent of the gradient entries. Parameter $q$ denotes the {\em tolerance},  which is a design parameter and can be chosen according to the learning problem. In the scope of this work, we utilize the random circularly shifted (RCS) code \cite{icassp-ext}, which allows workers to change codewords over time. In a broad sense, in RCS codes, $\mathbf{W}$ is divided into $K$ submatrices $\mathbf{W}_{1},\ldots,\mathbf{W}_{K}\in\mathbb{R}^{d/K\times d}$ and those submatrices are concatenated to form $\bar{\mathbf{W}}=[\mathbf{W}_{1},\ldots,\mathbf{W}_{K}]$. Then an assignment matrix, showing assigned submatrices to each worker, is formed by operating random circular shifts on $\bar{\mathbf{W}}$. Once the assignment matrix is established, codewords for each worker $i$ can be constructed by combining those submatrices in the $i$th column according to a certain order. 
 
 Next, we illustrate how RCS codes can be adapted to timely dynamic encoding.

\subsection{ Partial Recovery and Timely Dynamic Encoding}\label{dynamic_enc}
Dynamic encoding process consists of three phases namely; {\em data partition}, {\em ordering} and {\em encoding}, where the corresponding operators are denoted by $D(\cdot)$, $O(\cdot)$ and $E(\cdot)$, respectively.
Data partition operator distributes the submatrices $\mathcal{W}=\left\{\mathbf{W}_{1},\ldots,\mathbf{W}_{K}\right\}$ among the workers such that
\begin{equation}
D_{i}(\mathcal{W},M): \mathcal{W} \mapsto  \mathcal{W}_{i},~~ \vert \mathcal{W}_{i}\vert \leq M,
\end{equation}
where, $M$ is a given memory constraint and $\mathcal{W}_{i}$ is the set of assigned submatrices to the $i$th worker. We assume that operator $D(\cdot)$ is executed, for each worker, only once before the process, and thus  set $\mathcal{W}_{i}$ remains the same over the iterations.
The order operator $O(\cdot)$ is used to form an ordered set from the initial set $\mathcal{W}_{i}$ for encoding, i.e.,
\begin{equation}
O_{i,t}(\mathcal{W}_{i}): \mathcal{W}_{i} \mapsto  \tilde{\mathcal{W}}_{i,t}, 
\end{equation}
where $\tilde{\mathcal{W}}_{i,t}$ is an ordered set representing the order of computation at each iteration $t$ for the $i$th worker. We remark  that unlike the data partition operator, order operator may change over time. These two operators together can be represented by an assignment matrix $\mathbf{A}_{t}$, whose $i$th column is given by $\tilde{\mathcal{W}}_{i,t}$.

Once the assignment matrix $\mathbf{A}_{t}$ is fixed, the encoding process is executed according to a {\em degree vector} $\mathbf{m}$, which identifies the degree of each codeword based on its computation order. Encoding is executed for each worker independently. The encoder operator $E(\cdot)$ maps the ordered set of data (submatrices) to ordered set of codewords of size $L$, where $L$ is the length of $\mathbf{m}$, i.e.,
\begin{equation}
E_{i,t}(\tilde{\mathcal{W}}_{i,t},\mathbf{m}): \tilde{\mathcal{W}}_{i,t} \mapsto  \tilde{\mathcal{C}}_{i,t}=\left\{\mathbf{C}^{t}_{i,1},\ldots,\mathbf{C}^{t}_{i,L}\right\}.
\end{equation}

\begin{figure}[t]
	\centering  \includegraphics[width=0.9\columnwidth]{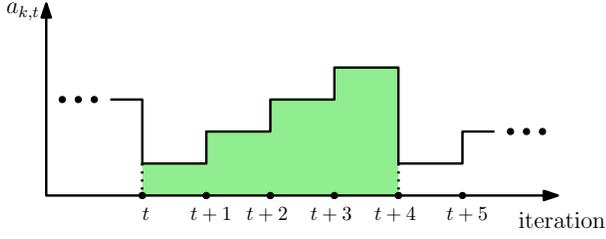}
	\caption{Sample age evolution of partial computation $\mathbf{W}_k\boldsymbol{\theta}$. Time $t$ marks the beginning of iteration $t+1$. $\mathbf{W}_k\boldsymbol{\theta}$ is recovered at iterations $t$ and $t+4$. }
	\label{fig:age_evol}
	\vspace{-0.5cm}
\end{figure}

The encoding operator first divides set $\tilde{\mathcal{W}}_{i,t}$ into $L$ disjoint subsets, $\mathcal{W}^{t}_{i,1},\ldots,\mathcal{W}^{t}_{i,L}$,  such that $\vert\mathcal{W}^{t}_{i,l}\vert=\mathbf{m}(l)$. Then, at iteration $t$, the coded submatrix of the $i$th worker with computation order $\ell$, denoted by $\mathbf{C}^{t}_{i,\ell}$, is constructed as
\begin{equation}
\mathbf{C}^{t}_{i,\ell}=\sum_{\mathbf{W}_{k}\in\mathcal{W}^{t}_{i,\ell}}\mathbf{W}_{k}.
\end{equation}

An example assignment matrix $\mathbf{A}_{t}$ is given below for $K=20$ and $M=6$:
 \begin{equation}
\mathbf{A}_{t}=
  \begin{bmatrix}
  {\color{blue}\mathbf{W}_{1}} & {\color{blue}\mathbf{W}_{2}} & {\color{blue}\mathbf{W}_{3}} & \dots  & {\color{blue}\mathbf{W}_{20}} \\
  {\color{red}\mathbf{W}_{4}}\ & {\color{red}\mathbf{W}_{5}} & {\color{red}\mathbf{W}_{6}} & \dots  & {\color{red}\mathbf{W}_{3}} \\
  {\color{red}\mathbf{W}_{11}} & {\color{red}\mathbf{W}_{12}} & {\color{red}\mathbf{W}_{13}} & \dots  & {\color{red}\mathbf{W}_{10}}\\
  {\color{brown}\mathbf{W}_{15}} & {\color{brown}\mathbf{W}_{16}} & {\color{brown}\mathbf{W}_{17}} & \dots  & {\color{brown}\mathbf{W}_{14}} \\
  {\color{brown}\mathbf{W}_{6}} & {\color{brown}\mathbf{W}_{7}} & {\color{brown}\mathbf{W}_{8}} & \dots  & {\color{brown}\mathbf{W}_{5}} \\
  {\color{brown}\mathbf{W}_{18}} & {\color{brown}\mathbf{W}_{19}} & {\color{brown}\mathbf{W}_{20}} & \dots  & {\color{brown}\mathbf{W}_{17}}\\
  \end{bmatrix}.\label{rcsc_dist}
 \end{equation}
 
The elements of the assignment matrix $\mathbf{A}_{t}$ are colored  to illustrate the first step of the encoding operator, for  $\mathbf{m}=[1, 2, 3]$, where colors blue, red and yellow represent the submatrices used to generate the first, second, and third codewords, respectively. The encoding phase for the first worker at iteration $t$ is illustrated below:
 \begin{equation}
\tilde{\mathcal{W}}_{1,t} =
  \begin{bmatrix}
  {\color{blue}\mathbf{W}_{1}} \\
  {\color{red}\mathbf{W}_{4}} \\
  {\color{red}\mathbf{W}_{11}} \\
  {\color{brown}\mathbf{W}_{15}} \\
  {\color{brown}\mathbf{W}_{6}} \\
  {\color{brown}\mathbf{W}_{18}} \\
  \end{bmatrix} \rightarrow
  \tilde{\mathcal{C}}_{1,t} =
  \begin{bmatrix}
   \mathbf{W}_{1} \\
   \mathbf{W}_{4} + \mathbf{W}_{11} \\
   \mathbf{W}_{15} + \mathbf{W}_{6} + \mathbf{W}_{18}\\
  \end{bmatrix}.\label{encoding_exp}
 \end{equation}
With this code, the worker first computes $\mathbf{W}_{1}\boldsymbol{\theta}_t$ and sends the result directly to the PS. Then, it computes $(\mathbf{W}_{4} + \mathbf{W}_{11})\boldsymbol{\theta}_t$ sends the result to the PS. Finally, it computes $(\mathbf{W}_{15} + \mathbf{W}_{6} + \mathbf{W}_{18})\boldsymbol{\theta}_t$ and sends the result to the PS.

Next, we formally state the problem using the data partition, ordering and encoding operators.

\subsection{Problem Definition}
The recovery of a partial computation $\mathbf{W}_k\boldsymbol{\theta}$ at iteration $t$ depends on the data partition $\left\{ D_i\right\}_{i\in[K]}$, ordering decisions $\left\{ O_{i,t}\right\}_{i\in[K]}$, encoding decisions $\left\{ E_{i,t}\right\}_{i\in[K]}$, computation delay statistics of the workers, $\mathbf{d}_t$, and the tolerance $q$, i.e.,
\begin{equation}
\mathbf{r}_t=R(D, O_t, E_t, d_t,q),
\end{equation}
where $R$ is the recovery operation that returns a vector $\mathbf{r}_t$ which demonstrates the recovered partial computations such that $\mathbf{r}_t(k) = 1$ if $\mathbf{W}_k\boldsymbol{\theta}_{t}$ is recovered at the PS for $k\in[K]$.\\
\indent In the partial recovery approach, without any further control on the assigned computations, operators are fixed throughout the training process. Thus, recovered submatrix indices may be correlated over time and some partial computations may not be recovered at all. We note that this kind of recovery behavior may lead to divergence especially when $q$ is large, since the updates become biased. Our goal is to introduce a dynamic approach for the coded computation/partial recovery procedure to regulate the recovery frequency of each partial computation. For this, we first introduce an age-based performance metric.

We define the age of partial computation $\mathbf{W}_k\boldsymbol{\theta}$ at iteration $t$, denoted by $a_{k,t}$, as the number of iterations  since the last time the PS recovered  $\mathbf{W}_k\boldsymbol{\theta}$. The age for each partial computation is updated at the end of each iteration in the following way
\begin{equation}\label{age_update}
    a_{k,t+1} =
    \begin{cases}
     a_{k,t}+1, & \text{if $\mathbf{r}_t(k) = 0$} \\
      1,        & \text{if $\mathbf{r}_t(k) = 1$}
    \end{cases}.
  \end{equation}
  

A sample age evolution of a partial computation is shown in Fig. \ref{fig:age_evol}. Here, partial computation $\mathbf{W}_k\boldsymbol{\theta}$ is recovered at iterations $t$ and $t+4$. The average age of the partial computation $\mathbf{W}_k\boldsymbol{\theta}$ over the training interval of $T$ iterations is
\begin{align}
    a_k = \frac{1}{T}\sum^{T}_{t=1} a_{k,t}.
\end{align}

In order to make sure that each submatrix contributes to the model update as equally as possible during the training period, our goal is to keep the age of each partial computation under a certain threshold  $a_{th}$. Thus, our objective is  
 \begin{equation}
 \min_{\Pi(D,O,E)} \frac{1}{T}\sum^{T}_{t=1}\frac{1}{K}\sum^{K}_{k=1}\mathbbm{1}_{\left\{a_{t,k}>a_{th}\right\}}, \label{obj_fcn}
 \end{equation}
where $\mathbbm{1}_{x}$ is the indicator function that returns 1 if $x$ holds, 0 otherwise. Here, $a_{th}$ is a design parameter that determines the desired freshness level for the partial computations and can be adjusted according to the learning problem. We note that the problem in (\ref{obj_fcn}) is over all data partitions, ordering and encoding policies, thereby is hard to optimally solve. Instead of solving (\ref{obj_fcn}) exactly, we introduce a timely dynamic ordering technique that can be used to regulate the recovery frequency of the partial computations.

\section{Solution Approach: Timely Dynamic Ordering}

In this section, we introduce timely dynamic ordering to better regulate the ages of partial computations and to avoid biased updates. We keep the data partition and encoding operators fixed and change only the ordering operator dynamically. This timely dynamic ordering is implemented by employing a vertical circular shift in the assignment matrix. With this, we essentially change the codewords and their computation order, which in turn, changes the recovered indices. 

We first employ fixed vertical shifts for dynamic ordering. Then, we will dynamically adjust the shift amount based on the ages of the partial computations. 

\subsection{Fixed Vertical Shifts}

In this code, which we call RCS-1, we employ one vertical shift for each worker at each iteration. That is, the order operator becomes
\begin{equation}\label{vertical_shift}
O_{i,t}(\mathcal{W}_{i}): \mathcal{W}_{i} \mapsto circshift(\mathcal{W}_{i}, mod(t,L)), 
\end{equation}
where $circshift$ is  the circular shift operator and $mod(x,y)$ is a modulo operator returning the remainder of $x/y$. By using vertical shifts, coded computations transmitted to the PS from a particular worker change over time to prioritize certain partial computations. For example, if worker $1$ employs the ordered set $\tilde{\mathcal{W}}_{1,t}$ and codewords $\tilde{\mathcal{C}}_{i,t}$ specified in (\ref{encoding_exp}) at iteration $t$, after applying one vertical shift, its computation order and codewords at iteration $t+1$, are given by
 \begin{equation}
  \tilde{\mathcal{W}}_{1,t+1}=
    \begin{bmatrix}
  {\color{blue}\mathbf{W}_{4}} \\
  {\color{red}\mathbf{W}_{11}} \\
  {\color{red}\mathbf{W}_{15}} \\
  {\color{brown}\mathbf{W}_{6}} \\
  {\color{brown}\mathbf{W}_{18}} \\
  {\color{brown}\mathbf{W}_{1}} \\
  \end{bmatrix} \rightarrow
   \tilde{\mathcal{C}}_{i,t+1}=
  \begin{bmatrix}
   \mathbf{W}_{4} \\
   \mathbf{W}_{11} +\mathbf{W}_{15} \\
   \mathbf{W}_{6} + \mathbf{W}_{18} + \mathbf{W}_{1}\\
  \end{bmatrix}.
 \end{equation}

Here, we see that, at iteration $t$, the worker prioritizes the computation of $\mathbf{W}_1\boldsymbol{\theta}$, while in the next iteration computation of $\mathbf{W}_4\boldsymbol{\theta}$ is prioritized. We note that, in this method, the shift amount is fixed to one shift at each iteration, and is independent of the ages of the partial computations. 

Next, we introduce an age-based vertical shift scheme to control the order of computations.

\subsection{Age-Based Vertical Shifts}

In this code, which we call RCS-adaptive,
we choose the vertical shift amount based on the current ages of the partial computations. That is, instead of shifting by $1$ at each iteration, the shift amount changes across iterations based on the ages of the partial computations. To effectively avoid biased updates, we focus on recovering the partial computations with the highest age at the current iteration, that is, the computations that have not been recovered in a while. 
In line with the problem in (\ref{obj_fcn}), we term the partial computations with age higher than the threshold $a_{th}$ as aged partial computations, which need to be recovered as soon as possible. To this end, a vertical shift amount is selected that places the maximum number of aged partial computations in the first position in the non-straggling workers' computation order so that they have a higher chance of recovery in the next iteration. In particular, to determine the shift amount in iteration $t+1$, the PS  considers the computation order at the workers that have returned at least one computation in the previous iteration and determines a shift which places maximum number of aged partial computations in the first order in these workers. Upon determining the shift amount, every worker's assignment matrix is shifted by that amount in the next iteration. For example, if the age-based shift amount is $3$ in iteration $t+1$, then the first user has
 \begin{equation}
  \tilde{\mathcal{W}}_{1,t+1}=
    \begin{bmatrix}
  {\color{blue}\mathbf{W}_{15}} \\
  {\color{red}\mathbf{W}_{6}} \\
  {\color{red}\mathbf{W}_{18}} \\
  {\color{brown}\mathbf{W}_{1}} \\
  {\color{brown}\mathbf{W}_{4}} \\
  {\color{brown}\mathbf{W}_{11}} \\
  \end{bmatrix} \rightarrow
   \tilde{\mathcal{C}}_{i,t+1}=
  \begin{bmatrix}
   \mathbf{W}_{15} \\
   \mathbf{W}_{6} +\mathbf{W}_{18} \\
   \mathbf{W}_{1} + \mathbf{W}_{4} + \mathbf{W}_{11}\\
  \end{bmatrix}.
 \end{equation}

Here, in iteration $t+1$, the first worker prioritizes the computation of $\mathbf{W}_{15}\boldsymbol{\theta}$.

\section{Numerical Results}  

In this section, we provide numerical results for comparing the proposed age-based partial computation scheme to alternative static schemes using a model-based scenario for computation latencies. For the simulations, we consider a linear regression problem over synthetically created training and test
datasets, as in \cite{CC.4}, of size of $2000$ and $400$, respectively. We also assume that the size of the model $d = 1000$ and the number of workers $K = 40$ while each worker can return $L=3$ computations with $\mathbf{m}=[1,2,3]$. A single simulation includes $T=400$ iterations. For all simulations, we use learning rate $\eta = 0.1$. To model the computation delays at the workers, we adopt the model in \cite{entropy}, and assume that the probability of completing $s$ computations at any worker, performing $s$ identical matrix-vector multiplications, by time $t$ is given by
\begin{equation}\label{exp_compute}
  F_s(t) \triangleq
  \begin{cases}
    1-e^{-\mu(\frac{t}{s}-\alpha)}, & \text{if $t \geq s\alpha$} \\
    0, & \text{otherwise}. 
  \end{cases}
\end{equation}

\begin{figure}[t]
	\centering  \includegraphics[width=0.7\columnwidth]{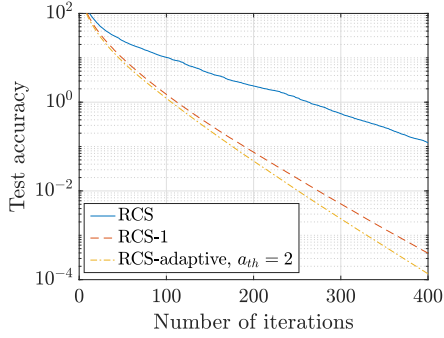}
	\caption{Test accuracy (log-scale) vs. number of iterations with $q=0.3$ and $15$ persistent stragglers.}
	\label{fig:1}
	\vspace{-0.2cm}
\end{figure}

First, we consider an extreme scenario in the straggling behavior, where we assume there are $15$ persistent stragglers that are fixed for all the $T=400$ iterations which do not complete any partial computations. For the non-persistent stragglers, we set $\mu = 10$ and $\alpha = 0.01$.\footnote{To simulate the straggling behavior in our simulations, we take $\alpha=10$ for the persistent stragglers so that effectively they do not complete any partial computations.} In Fig.~\ref{fig:1}, we set the tolerance level $q=0.3$, such that at each iteration the PS aims at recovering $28$ of the total $40$ partial computations. We see that the proposed timely dynamic encoding strategy with one vertical shift at each iteration, RCS$-1,$ achieves a significantly better convergence performance than the conventional static encoding with RCS. When the ages of partial computations are taken into consideration in determining the order of computation at each iteration with the proposed RCS-adaptive scheme with an age threshold of $a_{th} = 2$, we observe a further improvement in the convergence performance. 

\begin{figure}[t]
	\centering  \includegraphics[width=0.7\columnwidth]{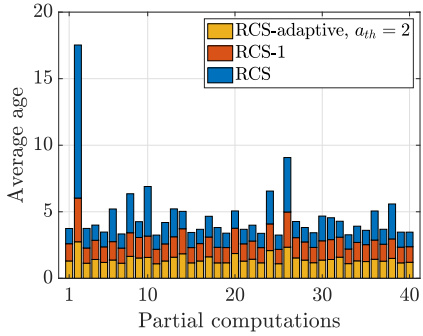}
	\caption{Average ages of the partial computations with $q=0.3$ and $15$ persistent stragglers.}
	\label{fig:2}
	\vspace{-0.2cm}
\end{figure}
\begin{figure}[t]
	\centering  \includegraphics[width=0.7\columnwidth]{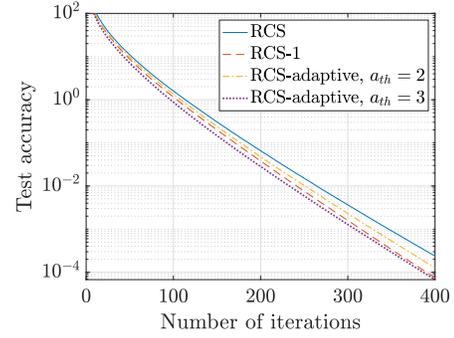}
	\caption{Test accuracy (log-scale) vs. number of iterations with $q=0.3$ and straggling behavior based on a 2-state Markov chain with a state transition probability of $p=0.05$.}
	\label{fig:14}
	\vspace{-0.2cm}
\end{figure}
\begin{figure}[t]
	\centering  \includegraphics[width=0.7\columnwidth]{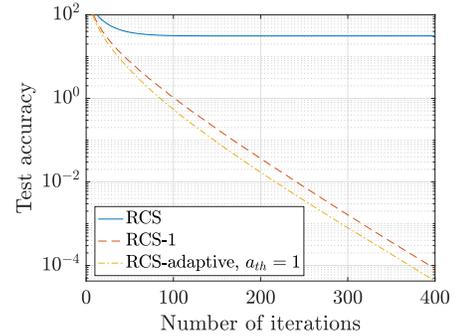}
	\caption{Test accuracy (log-scale) vs. number of iterations for the uncoded scheme with $q=0.2$ and $15$ persistent stragglers.}
	\label{fig:6.5}
	\vspace{-0.2cm}
\end{figure}

An interesting observation comes from Fig.~\ref{fig:2}, where we plot the average ages of the partial computations. While the proposed timely dynamic encoding strategy does not result in a better average age performance for every single partial computation, it targets the partial computations with the highest average age (see computation tasks $2$, $10$, and $26$ in Fig.~\ref{fig:2}). By utilizing the dynamic order operator, we essentially lower the average age of the partial computations with the worst age performance at the expense of slight increase in that of the some remaining partial computations. As expected, age-based vertical shift strategy further lowers the average ages of the partial computations. Here, we can draw parallels with this result and \cite{Jiang18d}, which shows that as long as each component is received every $p$ iterations, the distributed SGD can maintain its asymptotic convergence rate. From Fig.~\ref{fig:2}, we can see that the proposed vertical shift operator guarantees that on average each task is received every $3$ iterations, since the yellow bar in Fig.~\ref{fig:2} is less than $3$ for each partial computation. 

We note that in Figs.~\ref{fig:1} and~\ref{fig:2} the performance gap between RCS-1 and RCS-adaptive schemes is narrow. This shows that the randomness introduced by a fixed vertical shift is already quite helpful in mitigating the biased updates with less stale partial computations. 

In Table~\ref{table1}, we look at the value of the objective function in (\ref{obj_fcn}) when $a_{th}=2$ with $\mu = 10$ and $\alpha = 0.01$ for varying tolerance levels in the case of fixed $15$ persistent stragglers throughout all the iterations. We observe that, for each tolerance level $q$, when RCS-1 is employed, we achieve a better performance than the static RCS scheme, whereas the age-based vertical shift method RCS-adaptive results in the best performance. This is because the RCS-adaptive scheme specifically targets the computational tasks that have average age higher than the threshold $a_{th}$ to effectively create less biased model updates where each partial computation contributes to the learning task more uniformly. 
\begin{table}[t]
\centering
\vspace{2 mm}
\begin{tabular}{ |c|c|c|c|c|c| }
\hline
Tolerance level & RCS & RCS-1  & RCS-adaptive \\
\hline
$q=0.1$ & $0.0261$ & $0.0180$ & $0.0156$  \\ 
\hline
$q=0.2$& $0.0681$ & $0.0476$ & $0.0451$  \\ 
\hline 
$q=0.3$& $0.1316$ & $0.0970$ & $0.0919$ \\ 
\hline
\end{tabular} 
\vspace{1 mm}
\caption{ \label{table1} The value of the objective function in (\ref{obj_fcn}) when $a_{th} = 2$ for varying tolerance levels $q$.}
\vspace{-5 mm}
\end{table}

Second, we consider a more realistic scenario and model the straggling behavior of workers based on a two-state Markov chain: a slow state $s$ and a fast state $f$, such that computations are completed faster when a worker is in state $f$. This is similar to the Gilbert-Elliot service times considered in \cite{Yang19, Buyukates20a}. Specifically, in (\ref{exp_compute}) we have rate $\mu_f$ in state $f$ and rate $\mu_s$ in state $s$ where $\mu_f > \mu_s$. We assume that the state changes only occur at the beginning of each iteration with probability $p$; that is, with probability $1-p$ the state stays the same. A low switching probability $p$ indicates that the straggling behavior tends to stay the same in the next iteration. In Fig.~\ref{fig:14}, we set $p=0.05$, $q=0.3$, $\alpha = 0.01$, $\mu_S = 2$, and $\mu_f = 10$ and let $15$ workers start at the slow state, i.e., initially we have $15$ straggling workers. We note that with $15$ initial stragglers and $p=0$ we recover the setting considered in Fig.~\ref{fig:1}. We observe in Fig.~\ref{fig:14} that the proposed timely dynamic encoding strategy improves the convergence performance even though the performance improvement is less compared to the setting in Fig.~\ref{fig:1}. This is because, in this scenario, the straggling behavior is less correlated over iterations, which results in less biased model updates even for the static RCS scheme. Further, we see in Fig.~\ref{fig:14} that the RCS-adaptive scheme with $a_{th} =3$ performs the best, whereas the RCS-1 scheme outperforms the RCS-adaptive scheme when $a_{th} =2$. This shows that the age threshold $a_{th}$ needs to be tuned to get the best performance from the RCS-adaptive scheme.

Even though we focus on the distributed coded computation scenario in this work, the proposed dynamic order operator can be applied when the computations are assigned to workers in an uncoded fashion as well. To see the performance in the case of uncoded computations with MMC, we set $\mathbf{m}=[1,1,1]$ and $q=0.2$ and consider the same setup as in Fig.~\ref{fig:1}. In Fig.~\ref{fig:6.5}, we observe that the static partial recovery scheme fails to converge since if coding is not implemented along with partial recovery, model updates are highly biased in the presence of persistent stragglers. However, when the dynamic order operator is employed, particularly the age-aware vertical shifts with $a_{th}=1$, convergence is achieved. 



\section{Conclusion}
MMC and partial recovery are two strategies designed to enhance the performance of coded computation employed for straggler-aware distributed learning. The main drawback of the partial recovery strategy is biased model updates that are caused when the straggling behaviors of the workers are correlated over time. To prevent biased updates, we introduce a timely dynamic encoding strategy which changes the codewords and their computation order over time. We use an age metric to regulate the recovery frequencies of the partial computations. By conducting several experiments on a linear regression problem, we show that dynamic encoding, particularly an age-based encoding strategy, can significantly improve the convergence performance compared to conventional static encoding schemes. Although our main focus is on coded computation, the advantages of the proposed strategy are not limited to the coded computation scenario. The proposed timely dynamic encoding strategy can be utilized for coded communication and uncoded computation scenarios as well. 

\bibliographystyle{unsrt}
\bibliography{IEEEabrv,lib_v5,refs.bib}

\end{document}